\documentclass[aip,reprint,twocolumn,jcp,showpacs,floatfix]{revtex4-1}

\usepackage{dcolumn}
\usepackage{amsmath}

\usepackage{booktabs}
\usepackage{graphicx}
\usepackage{multirow}
\usepackage{color} % Allow text colors

\begin{document}

\title{Application of sampling theory in modeling of continuum processes: photoionization cross-sections of atoms.}
\author{A. Kozlov, S. Saha, H. M. Quiney}

\affiliation{ARC Centre of Excellence for Advanced Molecular Imaging}
\affiliation{School of Physics, The University of Melbourne, 
Victoria 3010, Australia}
\date{\today}

\begin{abstract}
We describe a method for the calculation of photoionization cross-sections using square-integrable amplitudes obtained from the diagonalization of finite-basis set representations of the electronic Hamiltonian. Three examples are considered: a model example in which the final state is a free particle, the hydrogen atom and neutral atomic sodium. The method exploits the Whittaker-Shannon-Kotel'nikov sampling theorem, which is widely used in digital signal sampling and reconstruction. The approach reproduces known data with very good accuracy and converges to the exact solution with increase of the basis set size.
\end{abstract} 
\pacs{PACS: }
\maketitle
%***************************************************************************

\section{Introduction}

Photoionization processes play an important role in chemical physics as a fundamental probe of  the structural and dynamical properties of many-body systems. Photoelectron spectroscopy is often applied to identify the electronic structure of materials in different fields as well as in different contexts. For example, photoelectron spectroscopy is used in astrophysics, aeronomy, radiation chemistry, environmental and atmospheric chemistry, metrology, surface science and catalysis and new material development in order to obtain detailed information about the energy levels and electronic wavefunctions of molecules and ions, as well as their interactions with radiation.

The development of theoretical models of electrodynamical processes in complex systems is driven primarily by the development in experimental techniques. Synchrotron radiation sources and the recent emergence of high-brightness x-ray free electron laser sources has stimulated renewed research activity in the area of molecular photoionization, which impacts the achievable resolution of methods of structure determination in molecules and clusters. A theoretical method suitable for the calculation of electrodynamical processes in molecules or clusters ultimately depends on the availability of the appropriate transition matrix elements involving continuous states, obtained either by detailed calculation or by some approximation scheme.

One of the unresolved issues in scattering theory in general, and photoionization theory in particular, involves the representation of the photoelectron wavefunction in a continuum state. The available theoretical methods for calculating the transition matrix can be broadly characterized as either employing a projection of the continuum wavefunction onto a square integrable ($L^2$) basis set or the construction of a true continuum wavefunction with appropriate boundary conditions \cite{Richards1984,Carravetta93b}. For molecules, the lack of spherical symmetry and the need to generate delocalized bound-state functions as initial states has led to a highly developed technology for generating electronic structures using $L^2$ basis sets, typically involving Gaussian functions. This technology is almost wholly independent of any method that attempts to construct a true scattering type solution, so performing all determinations of electrodynamic processes for molecules within an $L^2$ basis seems to be preferable if it can be made to be feasible and accurate.

The simplest model within the assumptions of strong orthogonality and the use of a single-centre expansion of the  photoelectron wavefunction to impose appropriate spherical boundary conditions far from the interacting system involves either the use of a plane or Coulomb wave representation. Some recent developments implement \cite{Gozem15,Oana07,Oana09}  this simple model along with correlated Dyson orbital, representing the initial and final state of the system. This method provides a reasonably accurate molecular photoinization cross sections of polyatomic systems. The exchange interaction of the photoelectron with the residual core in this model is accounted for by the use of a partial effective charge of the core. While this procedure is motivated  by physical intuition the selection of the effective charge parameters is largely empirical. 

%Stieltjes method

A mathematical technique that exploits $L^2$ wavefunctions is based on Stieltjes moment theory, which is not directly related to the explicit representation of the continuum state. In this method, the final state of the photoelectron wavefunction is sampled by the discretized representation of the complete spectrum. The discrete transition matrix elements obtained from these $L^2$ states are used to extract the oscillator strength density in the continuum. In this way, the Stieltjes method avoids the need for the solution of the scattering equations by extracting cross sections from spectral moments of an oscillator strength distribution. The discretized samples of the continuum states are the excited states calculated using any $L^2$ method. Any electronic structure method that generates a discretized representation of the complete spectrum can be used, in principle,  to calculate the photoionization cross section, of any other continuum process, using the Stieljes imaging method. This approach has been led by Langhoff and collaborators \cite{Langhoff1973a,Hermann1981,Langhoff1996,Langhoff79}. Further developments and applications of this method based on various \textit{ab-initio} electronic structure theory have been performed recently \cite{Ruberti2013,Cukras13,Ruberti2014,Ruberti2014a}. A number of other approaches based on $L^2$ discretizaton methods have also been reported, including the complex coordinate methods \cite{Rescigno1988a} and the reactance or K-matrix method \cite{Fano68}.

The accurate representation of the true continuum states in the field of residual molecular region can be achieved by allowing non-orthogonality of the photoelectron wavefunction \cite{Han2012} and exploiting variational techniques or by considering extra bound electron configurations in the wavefunctions, such as is implemented in R-matrix methods \cite{Tennyson2010}. The variational principle provides a powerful way of solving a wide range of differential and integral equations involved in  calculating collision and photoionization cross sections \cite{nesbet80book}.
The Kohn variational principle is probably the most widely used variational principle in describing collision phenomena \cite{Rescigno95}, though there are at least two major difficulties associated with implementations of this method. The first of these is the occurrence of anomalous singularities in the reactants or K-matrices that make these techniques difficult to apply in large scale calculations. The second problem concerns computational difficulties associated with the evaluation of multi-centre integrals involving both bound and continuum functions \cite{Schneider1986a,Schneider1989,Schneider1989a,Schneider90}. These issues have been addressed since the late 1980s and have been further developed into the complex Kohn method for electron-molecule scattering in recent years.

McKoy and co-workers \cite{McKoy1984,Takatsuka1981,Watson1979,Lucchese1980a,Lucchese1986,Lucchese1999} have developed methods that are based on, or in some way related to, the Schwinger variational principle or extensions of it that rely on the Lippmann-Schwinger scattering equation. The major disadvantage of this method is the occurrence of the Green`s function in this variational principle which is known in closed form only in some special cases, such as the free-particle case.  This method does have some advantages, however, including automatic incorporation of the correct boundary conditions.  The method then requires trial wave functions only in the region where the scattering potential is non-vanishing, allowing the exclusive use of discrete basis functions as trial functions in the solution. This feature is particularly desirable in applications to the  scattering of low-energy electrons by the nonspherical force fields of molecules and molecular ions.

In the present work, however, we have applied sampling theory methods originally developed for the signal processing and information theory to construct matrix elements involving continuum state amplitudes from discrete representations of the complete set of eigenfunctions of the electronic spectrum.  The calculations were carried out for hydrogen and sodium atoms. These systems are well-studied and are widely used as a benchmark for photoionization calculations. Another model, discussed in detail, involves a hydogenic initial state and a final state that is free particle. This model allows a simple analytic solution and may be used for illustrative purposes. The proposed approach significantly simplifies the computational labour and remains open to further development for application to molecular systems where sampling of a discrete representation of the complete spectrum is effectively mandatory.

%\annote[ss]{}{Need to add some more points on the novelty, strength and weakness of the new approach}

%\annote[ss]{}{moved Alex's intro to method section}
%}

\section{Theory}

The photoionization cross-section in the dipole approximation is given by  \cite{BetheSalpeterBook}
\begin{equation}\label{PhotoIon}
\sigma(\omega) = 2\pi \alpha \omega \sum_f |\left \langle i|\vec r \,|f\rangle \right|^2,
\end{equation}
where $\alpha$ is the fine structure constant, $\omega$ is the photon frequency, $| i \rangle$ is the initial state of the system, and  $\left|f\rangle\right.$ is the final state. Conservation of energy, $\hbar\omega = E_f-E_i$, is assumed to be fulfilled while the effect of recoil is ignored. In the case of atoms and molecules, the initial state in Eq. (\ref{PhotoIon}) is a bound state that can be calculated using number of well established methods. For simple systems, such as light atoms and molecules, K-matrix \cite{Cacelli1991, Cacelli2000} and R-matrix \cite{BurkeBook} methods can be employed to calculate the final state wavefunction, as well as convergent close-coupling \cite{Kheifets1998, Kheifets2009}. In general, however, these methods are limited by the rapid growth of computational complexity with the increase in the number of atoms and technical difficulties that arise through the loss of spherical symmetry. Hybrid basis set approaches have been proposed to tackle this problem \cite{Marante2014}, which aim to provide accurate representations of the continuum wavefunction within a finite box using B-spline functions and appropriate boundary conditions. This flexibility is achieved, however, at the expense of a very large basis set dimension and the need to calculate new types of interaction integral, which complicates its implementation within existing quantum chemistry packages.

Rather than attempting to construct an accurate representation of a molecular continuum state wavefunction, it seems preferable to deduce the values of matrix elements involved in Eq. (\ref{PhotoIon}) by some other means that is more consistent with the $L^2$ methods that are used to generate the target bound-states. As will be shown below, this can be achieved with the help of conventional $L^2$ Gaussian basis set that is widely used in the molecular structure calculations that form the practical basis of much of quantum chemistry. Such calculations are of particular importance for investigation of photoionization in medium to large size molecules which takes place in a number of applications such as the radiation damage processes that influence the quality of single molecule imaging algorithms \cite{Quiney2011}. 

Briefly, these $L^2$ methods solve the generalized matrix eigenvalue equations
\begin{equation}\label{DiagHam}
%\sum_{\mu =1}^N \langle \mu | \hat H | \nu \rangle C_{\epsilon \mu} = \sum_{\mu = 1}^N E_{\nu}\langle \mu | \nu \rangle C_{\epsilon \mu},
\mathbf{H}\mathbf{c}_{\epsilon} = E_{\epsilon}\mathbf{S}\mathbf{c}_{\epsilon}
\end{equation}
where $H_{\mu \nu}=\langle \mu | \hat H | \nu\rangle$ is a matrix element of Hamiltonian, $\hat{H}$ of the system, $| \mu \rangle$ and $| \nu \rangle$ are the basis states, $c_{\mu \epsilon}$ is an element of the eigenvector, $\mathbf{c}_{\epsilon}$, $E_{\epsilon}$ is an eigenvalue and $S_{\mu\nu} = \langle\mu| \nu\rangle$ is an overlap matrix elements involving $|\mu \rangle$ and $ \nu\rangle$. The eigenstates of the Hamiltonian are $|\epsilon\rangle = \sum_{\mu} c_{\mu, \epsilon}|\mu\rangle$. Within the Hartree-Fock approximation, the system of equations defined by (\ref{DiagHam}) is referred to as the Hartree-Fock-Roothaan equations or, simply, the Roothaan equations \cite{Roothaan1951}.
 
Use of $L^2$ discretized ``pseudo-states" as a basis to extract information about continuum processes is not a new idea. For example Heller, Reinhardt, and Yamani \cite{Heller1973} applied this method using a matrix representation of the free-particle and hydrogenic Hamiltonians in Laguerre-Sturmian basis. In these cases, basis set matrix diagonalization can be carried out analytically, allowing a comprehensive study to be performed. Here, we generalize their approach in a manner that does not rely on a particular choice of basis set but which does require, as is always the case in molecular electronic structure calculations, a numerical diagonalization of the matrix representation of the Hamiltonian.

Suppose one needs to estimate a matrix element $\langle i|\hat A|k\rangle$ of some operator $\hat A$ for the system of interest, for example an atom or a molecule. The final state of the photoelectron, $|k\rangle$, has positive energy, so that 
\begin{align}
& \hat H |k\rangle = \frac{k^2}{2}|k\rangle,\label{HamCont}\\
& \langle k'|k\rangle = \delta(k'-k),\label{ComplCont}
\end{align}
where $\delta(x)$ is the Dirac delta function, $k$ is the wavenumber and $k^2/2 = E > 0$ is the asymptotic kinetic energy of the continuum state. In addition, we assume that (\ref{DiagHam}) has been solved in some $L^2$ finite basis set, yielding the solutions 
\begin{align}
& \hat H |m\rangle = E_m |m\rangle,\label{HamPseudo}\\
& \langle n|m\rangle = \delta_{nm},\label{ComplPseudo}
\end{align}
where $\delta_{nm}$ is Kronecker delta function.  The continuum states $|m\rangle$ are identified as the ``positive-energy" solutions of the above equation, for which $E_m = k_m^2/2$. Functions $|m\rangle$ are not the actual eigenstates $|k\rangle$, but rather some wavepacket representation of them, so we shall refer to these discretized positive energy states as ``pseudo-states". The identity (\ref{ComplPseudo}) can then be rewritten as 
\begin{equation}\label{DiracKronecher}
 \int\limits_0^{\infty}\langle n|k\rangle\langle k|m\rangle dk \approx \delta_{nm}.
\end{equation}
In the case of the free-particle Hamiltonian the states $|k\rangle$ form a complete set and Eq.(\ref{DiracKronecher}) is a strict equality. In other cases, however, the partitioning of the spectrum into ``positive" and ``negative" energy parts renders (\ref{DiracKronecher}) an approximation. It proves to be, however, an excellent approximation if $|m\rangle$ and $|n\rangle$ are discrete states that are classified, on the basis of their respective energies, as being of positive-energy type. The integral in (\ref{DiracKronecher}) can then be approximated using
\begin{align}\label{3integralAppr}
&  \int\limits_0^{\infty}\langle n|k\rangle \langle k| m\rangle dk \approx \sum\limits_{l=1}^{N}\langle n|k_l\rangle\langle k_l| m\rangle \omega_k^{(l)},\\
& \omega_k^{(l)} =  \left.\frac{dk}{ds}\right|_{s=l},
\end{align}
where $s$  $\epsilon$ $[1, N]$. For integer values $s = m$, the interpolation $k(s)$ consequently takes values  $k(m)=k_m$ that correspond to positive-energy solutions of (\ref{HamPseudo}). Substituting Eq. (\ref{3integralAppr}) into Eq. (\ref{DiracKronecher}) we arrive at the system of equations

%\begin{equation}\label{1integralAppr}
 %\int\limits_0^{\infty}\langle n|k\rangle \langle k| m\rangle dk \approx \int\limits_{k_1}^{k_N} \langle n|k\rangle\langle k| m\rangle dk,
%\end{equation}
%where wavenumber $k_1$ corresponds to the continuum eigenstate $|k_1\rangle$ and the first positive energy pseudo-state $|1\rangle$, while $k_N$ corresponds to $|k_N\rangle$ and the highest-energy pseudo-state, $|N\rangle$. To estimate the integral numerically on the right of (\ref{1integralAppr}), it is convenient to change the integration variable, so that 
%\begin{equation}\label{2integralAppr}
 %\int\limits_{k_1}^{k_N} \langle n|k\rangle\langle k| m\rangle dk = \int\limits_{1}^{N} \langle n|k(s)\rangle\langle k(s)| m\rangle \frac{dk}{ds} ds,
%\end{equation}
%where $s$  $\epsilon$ $[1, N]$. For integer values $s = m$, the interpolation $k(s)$ consequently takes values  $k(m)=k_m$ that correspond to positive-energy solutions of (\ref{HamPseudo}). The integral on the right of (\ref{2integralAppr}) can be approximated by a finite summation of the form
%\begin{align}\label{3integralAppr}
%& \int\limits_{1}^{N} \langle n|k(s)\rangle\langle k(s)| m\rangle \frac{dk}{ds} ds \approx \sum\limits_{l=1}^{N}\langle n|k_l\rangle\langle k_l| m\rangle \omega_k^{(l)},\\
%& \omega_k^{(l)} =  \left.\frac{dk}{ds}\right|_{s=l},
%\end{align}
%where the integral involving $ds$ has been evaluated using a simple quadrature by setting $\Delta s = 1$. Substituting (\ref{3integralAppr}), (\ref{2integralAppr}), and (\ref{1integralAppr}) into (\ref{DiracKronecher}) we arrive at the system of equations

\begin{equation}
\sum\limits_{l=1}^{N} \langle n|k_l\rangle\langle k_l| m\rangle \omega_k^{(l)} =\delta_{nm},
\end{equation}
where $\langle n|k_l\rangle$ are unknowns. The solutions of these equations are of the form
\begin{equation}\label{SFnodes}
\langle n|k_l\rangle = \frac{ \delta_{nl}}{\sqrt{\omega_k^{(l)}}} = \delta_{nl}\left.\sqrt{\frac{ds}{dk}}\right|_{k=k_l}.
\end{equation} 
Equation (\ref{SFnodes}) allows one to evaluate the matrix element $\langle i|\hat A|k_n\rangle$ if one knows the value of $\langle i|\hat A|n\rangle$:
\begin{equation}\label{HellerGen}
\langle i|\hat A|k_m\rangle = \sum\limits_{n=1}^N \langle i|\hat A|n\rangle\langle n|k_m\rangle=\langle i|\hat A|m\rangle\left.\sqrt{\frac{ds}{dk}}\right|_{k=k_m},
\end{equation}
where $s = s(k)$ is a smooth interpolation on the points $\{k_n, n\}$. The above result generalizes a method described by \citet{Heller1973} for a discrete set of wavepackets $\{|n\rangle\}$ obtained by discretizing a Hamiltonian in a finite basis set. The quality of the result obtained using (\ref{HellerGen}) depends on the completeness of the basis set $\{|n\rangle\}$, which was established in the particular case of a complete Laguerre-Sturmian basis in Refs.~\onlinecite{Heller1973,Yamani1975}.

\begin{figure}[t]
\center{\includegraphics[width=1\linewidth]{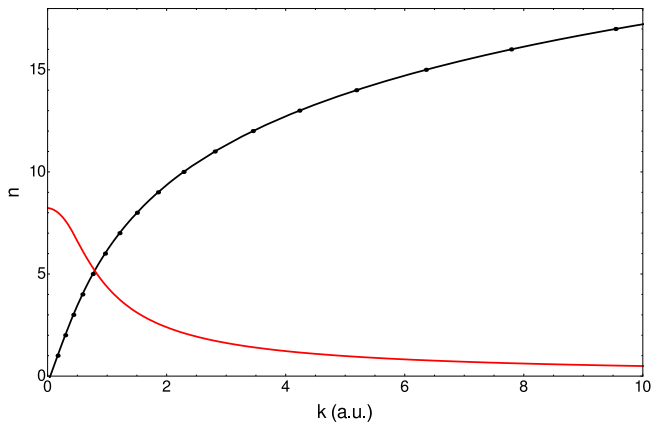} }
 \caption{Interpolation $s = s(k)$ (black line) of points $\{k_n, n\}$ (black dots) used for construction of sampling function (\ref{SFfree}) for free particle final state. Red line represents the derivative $ds/dk$. Values of $k_n$ were obtained by diagonalizing free particle Hamiltonian with 30 p-type Gaussian functions.}
\label{DensityFree}
\end{figure}

For the application of photoionization processes in (\ref{HellerGen}) we take $\hat A = \hat E1$, where $\hat {E1}$ is the electric dipole transition operator. For bound-continuum electric dipole matrix elements and the transition from a bound-state to a continuum pseudo-state of corresponding energy, we have   
\begin{align}\label{HellerE1}
& \langle i|\hat {E1}|k_m\rangle = \langle i|\hat {E1}|m\rangle\left.\sqrt{\frac{ds}{dk}}\right|_{k=k_m},\\
& \left.\langle i|\hat {E1}|k\rangle\right|_{k=k_m} \equiv \langle i|\hat {E1}|k_m\rangle
\end{align}
where $k_m$ is the wavenumber corresponding to the eigenvalue of pseudo-state $|m\rangle$, $s=s(k)$ is a smooth interpolation on the points $\{k_m, m\}$ and $|i\rangle$ is the initial bound state of the atom or molecule. Note that $ds/dk|_{k=k_n}\Delta k$ can be interpreted as the number of pseudo-states in the interval $\Delta k$ in the vicinity of $k = k_n$ (see for example Fig. \ref{DensityFree}) and that, therefore, $ds/dk|_{k=k_n}$ represents the density of pseudo-states in the vicinity of $k_n$. If the square-integrable basis set tends towards completeness, the density of pseudo-states tends to infinity for any wavenumber $k$.  

In order to calculate $\langle i|\hat {E1}|k\rangle$ for all values of $k$ and not just those values equal to the wavenumbers $k_m$ of pseudo-states, one can use a simple spline or other kind of polynomial interpolation. A more powerful way of calculating $\langle i|\hat {E1}|k\rangle$ within the approximations that have been imployed uses the Kramer sampling theorem \cite{Kramer1957}. Within this formalism, the matrix element for electric dipole transitions  can be written as
\begin{equation}\label{IntTransf}
\langle i|\hat {E1}|k\rangle = \lim_{N\rightarrow \infty} \sum_{m=1}^N \langle i|\hat {E1}|m\rangle \langle m| k\rangle,
\end{equation}
where the completeness condition $\sum |m \rangle \langle m| = 1$ has been used. Using (\ref{HellerE1}) the above equation can be written as 
\begin{align}
& \langle i|\hat {E1}|k\rangle = \lim_{N\rightarrow \infty} \sum_{m=1}^N \langle i|\hat {E1}|k_m\rangle S(k, k_m),\label{KramerT} \\
& S(k, k_m) = \langle m| k\rangle \left/ \sqrt{\frac{ds}{dk}}\right|_{k=k_m},\label{SFdef}
\end{align}
\\
where $s = s(k)$ is defined as in (\ref{HellerE1}). We shall refer to the functions $S(k, k_m)$ as ``Kramer sampling functions" and to the functions $\langle m | k\rangle$, which are the projections of pseudo-states onto the continuum states, as ``sampling functions". The Kramer sampling theorem formulated in the form of (\ref{KramerT}) allows one to calculate the matrix elements $\langle i|\hat {E1}|k\rangle$ for any value of the wavenumber $k$ if one knows the values of matrix elements for some discrete set of points ${k_m}$ called sampling points. In a particular case when the sampling points are equally spaced $s(k) = k$ and the continuum states are free-particle spherical outgoing s-type waves confined in the spherical box of radius $r_{max} = 2\pi$, the Kramer sampling formula reduces to well-known Whittaker-Shannon-Kotel'nikov sampling theorem and $S(k, k_m) \sim \text{sinc}(\pi k - \pi k_n)$.

In the general case, the functions $S(k, k_m)$ are unknown. In the limit of a complete $L^2$ basis functions, the functions defined in (\ref{SFdef}) satisfy the conditions
\begin{align}
& S(k_n, k_m) = \delta_{nm} \label{SFconditions1}\\
& \int S^*(k, k_m)S(k, k_n) \left.\frac{ds}{dk}\right|_{k=k_n} dk = \delta_{nm}.\label{SFconditions2}
\end{align}
Some examples of these functions, their properties, as well as more rigorous and detailed formulation of Kramer sampling theorem can be found in Refs.~\onlinecite{ZayedBook, Higgins1972}. It is shown in Ref.~\onlinecite{ZayedBook} that the replacement of the exact functions $S(k, k_m)$ given by (\ref{SFdef}) by functions which correspond to a different set of continuum states, leads to the same result in the limit of the infinite number of sampling points. In the present case we restrict our focus to atoms, so it is more convenient to use sampling functions derived from Bessel-Hankel transform \cite{Higgins1972} given by
\begin{equation}\label{SFmodel}
\tilde S(y, y_m) = \frac{2(y y_m)^{1/2}J_{\nu}(y)}{J'_{\nu}(y_m)(y^2-y_m^2)},
\end{equation}    
where $y_n$ are solutions of $J_{\nu}(y)=0$. To construct suitable sampling function from (\ref{SFmodel}) one can use the relation
\begin{equation}\label{SFmodelBad}
S(k, k_m) =\tilde S(y(k), y_m)\sqrt{\frac{dy}{dk}} \left/ \sqrt{\frac{dy}{dk}}\right|_{k=k_m},
\end{equation}   
where $y = y(k)$ is a smooth interpolation of points $\{k_n, y_n\}$. This function can be shown to satisfy (\ref{SFconditions1}), but in general, it doesn't satisfy (\ref{SFconditions2}). In the above equation function the limit of $dy/dk$ is $\pi ds/dk$ for $k \gg k_1$. For small wavenumbers, $dy/dk$ depends strongly on the interpolation procedure employed and varies dramatically between different orders of Bessel function in (\ref{SFmodel}). In calculations it is more convenient to construct sampling functions using
\begin{equation}\label{SFmodelGood}
S(k, k_m) =\tilde S(y(k), y_m)\sqrt{\frac{ds}{dk}}\left/ \sqrt{\frac{ds}{dk}}\right|_{k=k_m},
\end{equation}   
where, as above, $y = y(k)$ and $s = s(k)$ provide smooth interpolations of the points $\{k_n, y_n\}$ and $\{k_n, n\}$, respectively. For the function defined by (\ref{SFmodelGood}), condition (\ref{SFconditions1}) is satisfied exactly, while (\ref{SFconditions2}) holds only approximately, though to good accuracy,  according to our numerical calculations. Summarizing these results, Eq. (\ref{KramerT}, \ref{SFdef}, \ref{SFmodel}, \ref{SFmodelGood}) may be used to construct an approximation to the matrix element of the electric dipole operator according to 
\begin{align}
& \langle i|\hat {E1}|k\rangle = \sum\limits_{n=1}^N \langle i|\hat {E1}|n\rangle \langle n|k\rangle \label{IntpForm},\\
& \langle n|k\rangle = \frac{2(y(k)y_n)^{1/2}J_{\nu}(y(k))}{J'_{\nu}(y_n)(y^2(k)-y_n^2)} \sqrt{\frac{ds}{dk}}\label{SFfree}.
\end{align}
We denote the function $\langle n|k\rangle$ in (\ref{SFfree}) a ``Bessel-Hankel sampling function'', to reflect its relation to the Kramer sampling function $S(k, k_m)$ and the Bessel-Hankel transformation from which it is derived. The sign of $\langle n|k\rangle$ is undetermined, so for later convenience we impose the convention that $\langle r| n\rangle > 0$ for $r\to 0$. 

%\section{Sampling functions}
\section{Free particle photoionization}

It is convenient to start our analysis of photoionization using a simple model, in which initial state is a ground-state of hydrogen atom and final state is a free particle. This model has being discussed in Ref.~\onlinecite{LandauBook3} and treats potential energy as a perturbation. Its advantage is that most of the results can be obtained analytically, allowing direct comparison with our numerical approach. In the free-particle basis, the reduced electric-dipole transition matrix element is
\begin{equation}\label{AnalytFree}
\langle 1s\|r\|k p_{fp}\rangle =16\sqrt{ \frac{2}{\pi}}  \frac{k^2}{(1+k^2)^3},
\end{equation}
where $|1s\rangle$ represents the ground state of hydrogen atom and $\left|k p_{fp}\rangle \right.$ is p-type free particle final state with wavenumber $k = \sqrt{2E_f}$. At the same time, one can apply Eq. (\ref{IntpForm}) to calculate the matrix element (\ref{AnalytFree}) with functions $\left|m\rangle\right.$ obtained via diagonalization of the free particle Hamiltonian $\hat H_{fp} = -\nabla^2/2$ using p-type Gaussian basis functions
\begin{align}
& \langle 1s\|r\|k p_{fp}\rangle = \sum_m \langle 1s\|r\|m\rangle \langle m | k p_{fp}\rangle,\label{MEdecomp}\\
& \left\langle r |m\rangle\right. = \sum_{a=1}^N C_{ma} \frac{2^{\frac{11}{4}} \alpha_a^{\frac{5}{4}}}{\sqrt{3}\pi^{\frac{1}{4}}} r \text{exp}[-\alpha_a r^2]\label{FreeExp}
\end{align}
where $k_m = \sqrt{2E_m}$, and the expansion coefficients, $C_{ma}$, are obtained by solving the Roothaan equations, (\ref{DiagHam}), in the $p$-type Gaussian basis set. A total of $N=30$ basis set exponents with $\alpha_a = \alpha \beta^{a-1}$, $\alpha = 0.01$, and $\beta = 1.5$ were used; Convergence and linear dependence properties of this basis set are discussed in Ref.~\onlinecite{WilsonBook}. Note that in the case of a free particle all energies $E_m$ are positive in the final state. Since we know the exact solution for the final state wavefunction it is possible to calculate the sampling functions $\langle m | k p_{fp}\rangle$, using (\ref{FreeExp})
\begin{multline}\label{AnalytSF}
\langle m | k p_{fp}\rangle = \int\limits_0^{\infty} dr \langle m | r \rangle r\sqrt{kr} J_{3/2}(kr) = \\
 \sum_{a=1}^N C_{ma} \left(\frac{2}{\pi}\right)^{\frac{1}{4}}\frac{k^2}{\sqrt{3}\alpha_a^{\frac{5}{4}}}e^{-\frac{k^2}{4\alpha_a}}.
\end{multline}
\begin{figure*}[t]
\begin{minipage}[t]{0.49\linewidth}
\center{\includegraphics[width=1\linewidth]{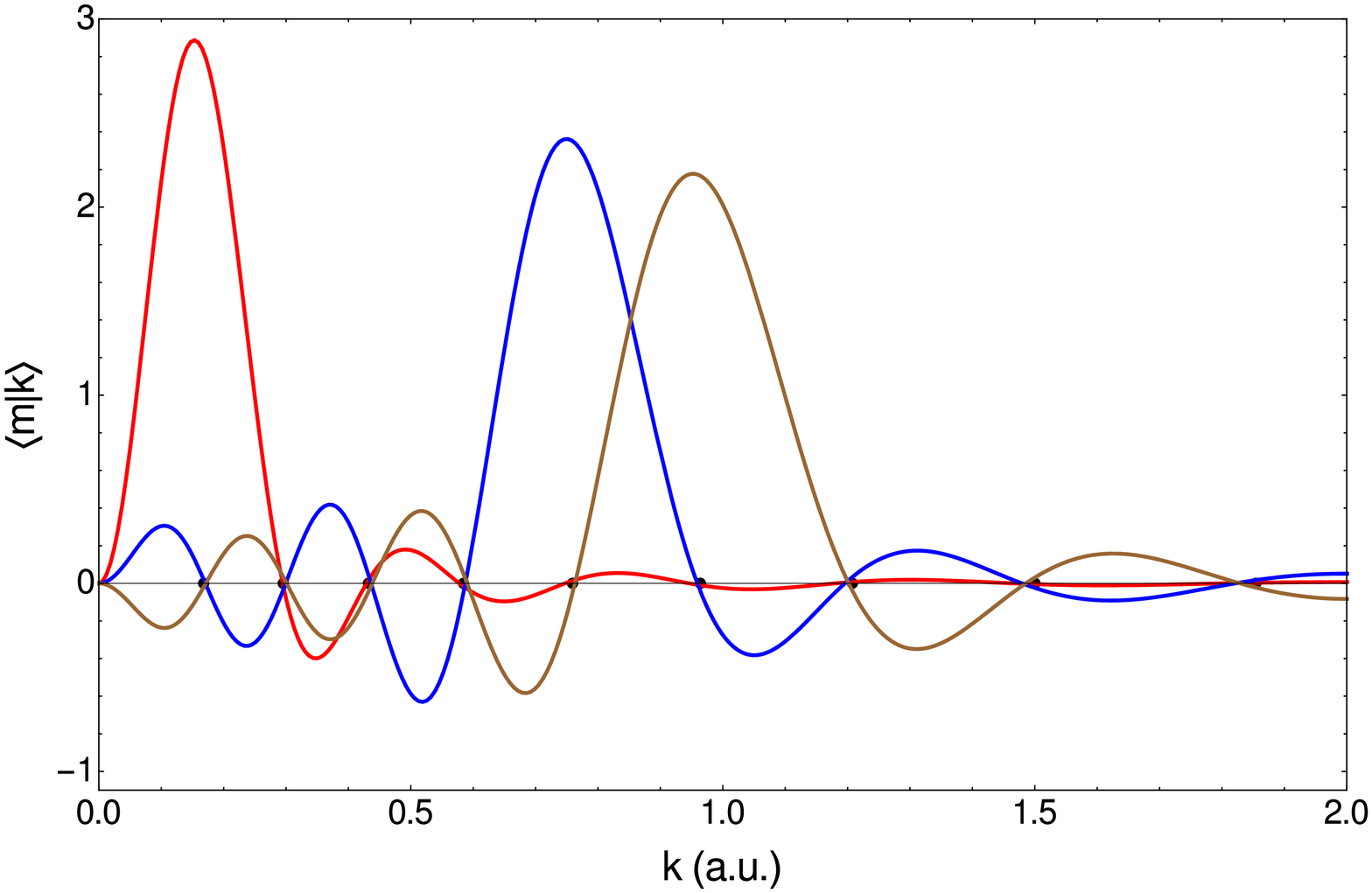} }
 \caption{ Sampling functions $\langle m | k\rangle$ given by (\ref{AnalytSF}). Solid lines represent the following functions:  $\langle 1 | kp_{fp}\rangle$ (red), $\langle 5 | kp_{fp}\rangle$ (blue), $\langle 6 | kp_{fp}\rangle$ (brown).  Solid black dots are the eigenvalues $k_m = \sqrt{2E_m}$ obtained from diagonalization of the free particle Hamiltonian. }
\label{FreeSF}
\end{minipage}
\hfill
\begin{minipage}[t]{0.49\linewidth}
\center{\includegraphics[width=1\linewidth]{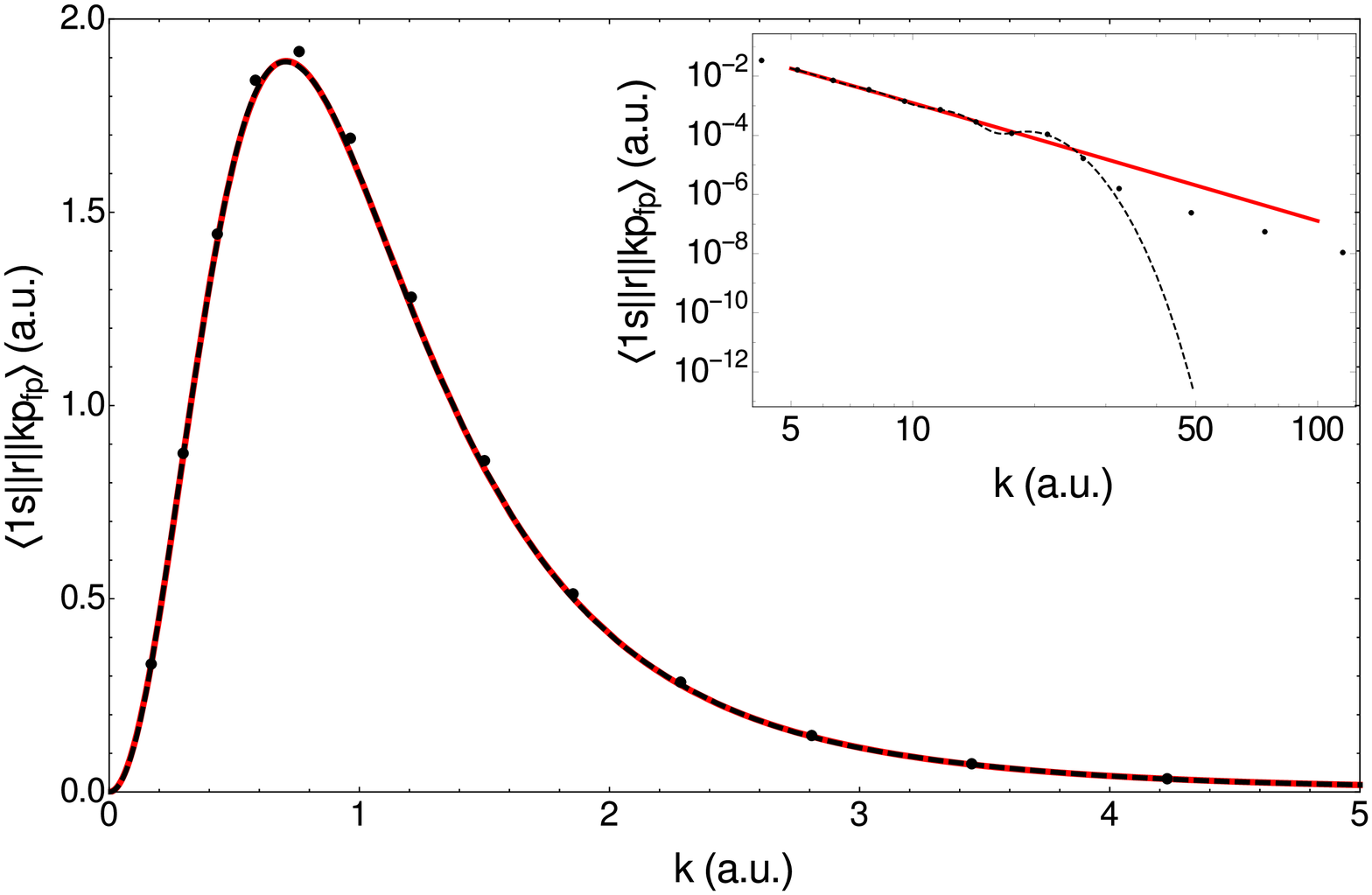} }
 \caption{Matrix element for transition from hydrogen ground state to spherical plane wave. Solid line represents analytic formula (\ref{AnalytFree}), dashed line represents numerical result using Gaussians (\ref{MEdecomp}). Solid black dots represent matrix elements with $k=k_m$ given by (\ref{MEdecomp}) with a single term $\langle m | k_m p_{fp}\rangle$ included.}
\label{Free}
\end{minipage}
\end{figure*}
Some of the sampling functions (\ref{AnalytSF}) are shown in Fig \ref{FreeSF}.  For the sake of brevity we use a simplified notation, so that $\langle m | k p_{fp}\rangle \equiv \langle m | k\rangle $. Substitution of sampling functions (\ref{AnalytSF}) in (\ref{MEdecomp}) leads to the representation of the matrix elements using discrete positive energy solutions. The results are presented in Fig. \ref{Free}. Note that the exact free particle continuum states were used to calculate sampling functions. As one can see the agreement between matrix elements calculated using Eq. (\ref{AnalytFree}) and Eq. (\ref{MEdecomp}) is quite good up to $k\approx 10$  a.u., which corresponds to energies of about $E_f = 600$ eV. For higher values of $k$, the interpolation procedure is unsatisfactory and inaccurate. This reveals a fundamental limitation about the particular basis set, rather than the use of $L^2$ amplitudes to extract information about continuum processes. In order to expand a rapidly oscillating continuum function using pseudo-states (\ref{FreeExp}) one needs a very dense set of $\alpha_a$ corresponding to $\beta\to 1$, but that leads to a well known problem of 
the linear dependence of the basis \cite{WilsonBook}. While this problem may be circumvented entirely using the Laguerre-Sturmian basis, the Gaussian basis is intrinsically limited in this regard. 
\begin{figure*}[t]
\begin{minipage}[t]{0.49\linewidth}
\center{\includegraphics[width=1\linewidth]{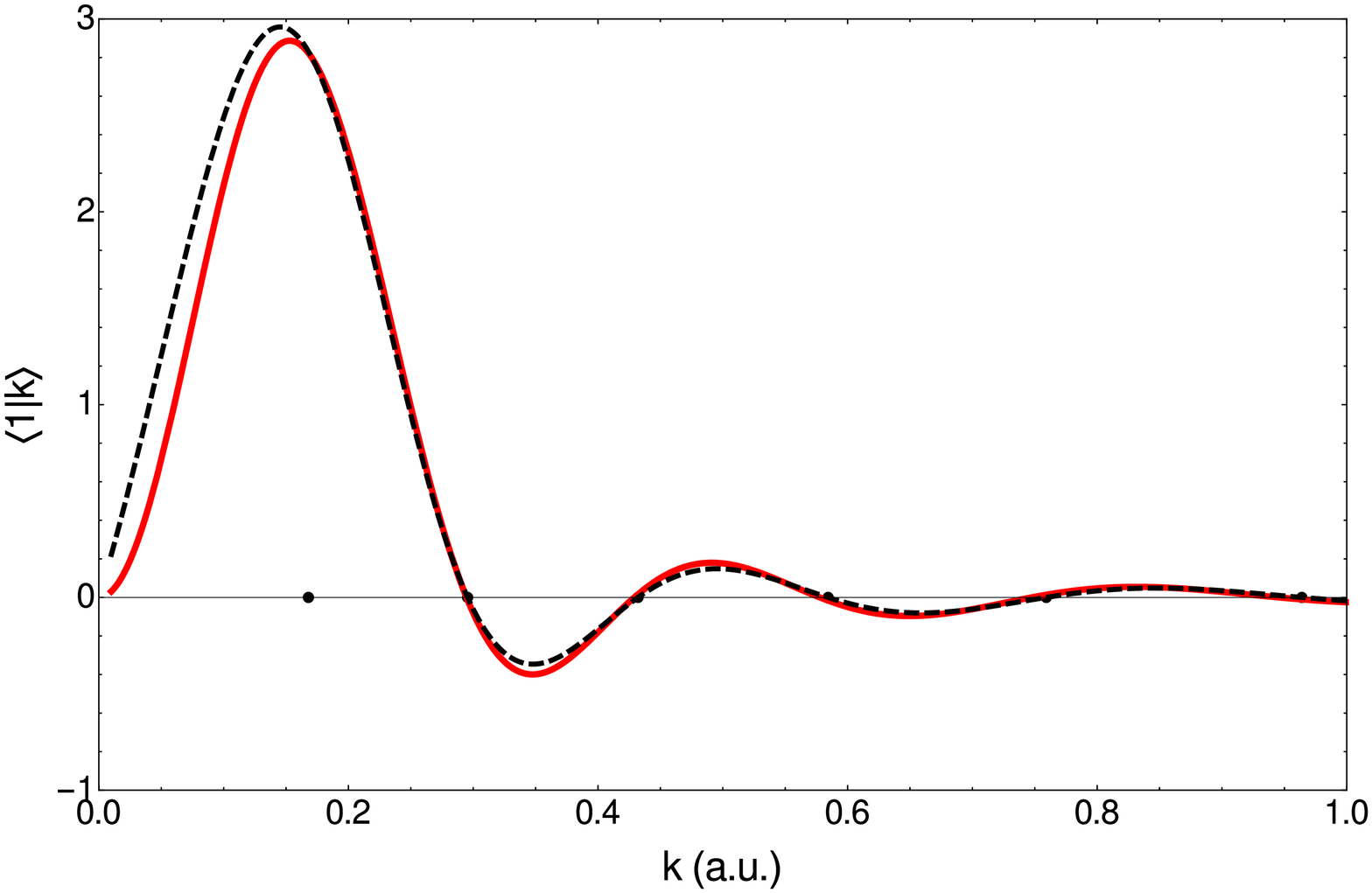} a) }
\center{\includegraphics[width=1\linewidth]{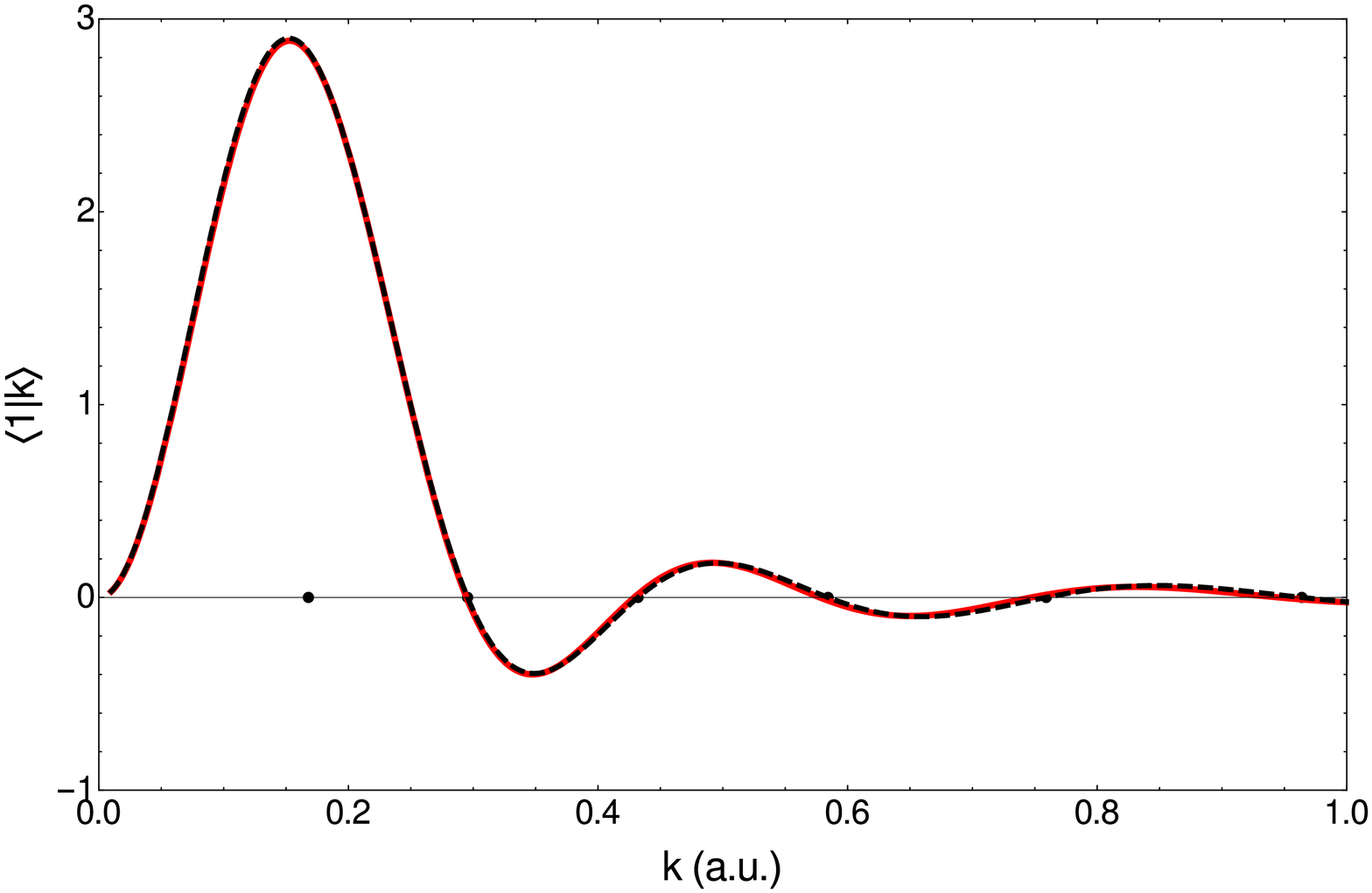} b) }
\center{\includegraphics[width=1\linewidth]{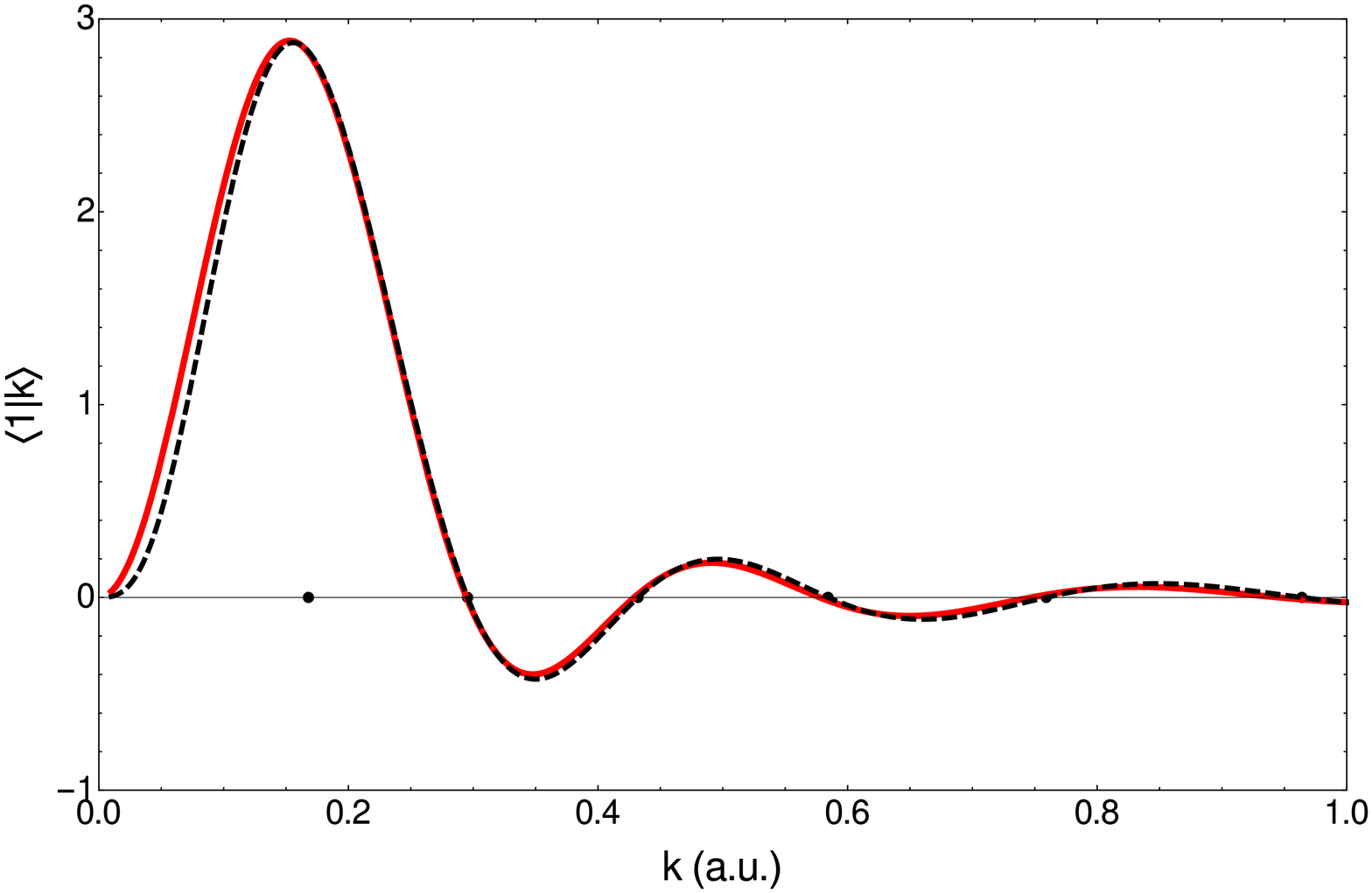} c) }
\caption{Comparison of ``exact" sampling function $\langle 1 | k\rangle$ given by the Eq. (\ref{AnalytSF}) and Bessel-Hankel sampling functions (\ref{SFfree}) with a) $\nu = 1/2$, b) $3/2$, c) $5/2$ from top to bottom respectively. Solid red line represents function given by the Eq. (\ref{AnalytSF}), dashed black line is Bessel-Hankel sampling function  (\ref{SFfree}). Solid black dots are the eigenvalues $k_m = \sqrt{2E_m}$ obtained from diagonalization of the free particle Hamiltonian. }
\label{SFcompare}
\end{minipage}
\hfill
\begin{minipage}[t]{0.49\linewidth}
\center{\includegraphics[width=1\linewidth]{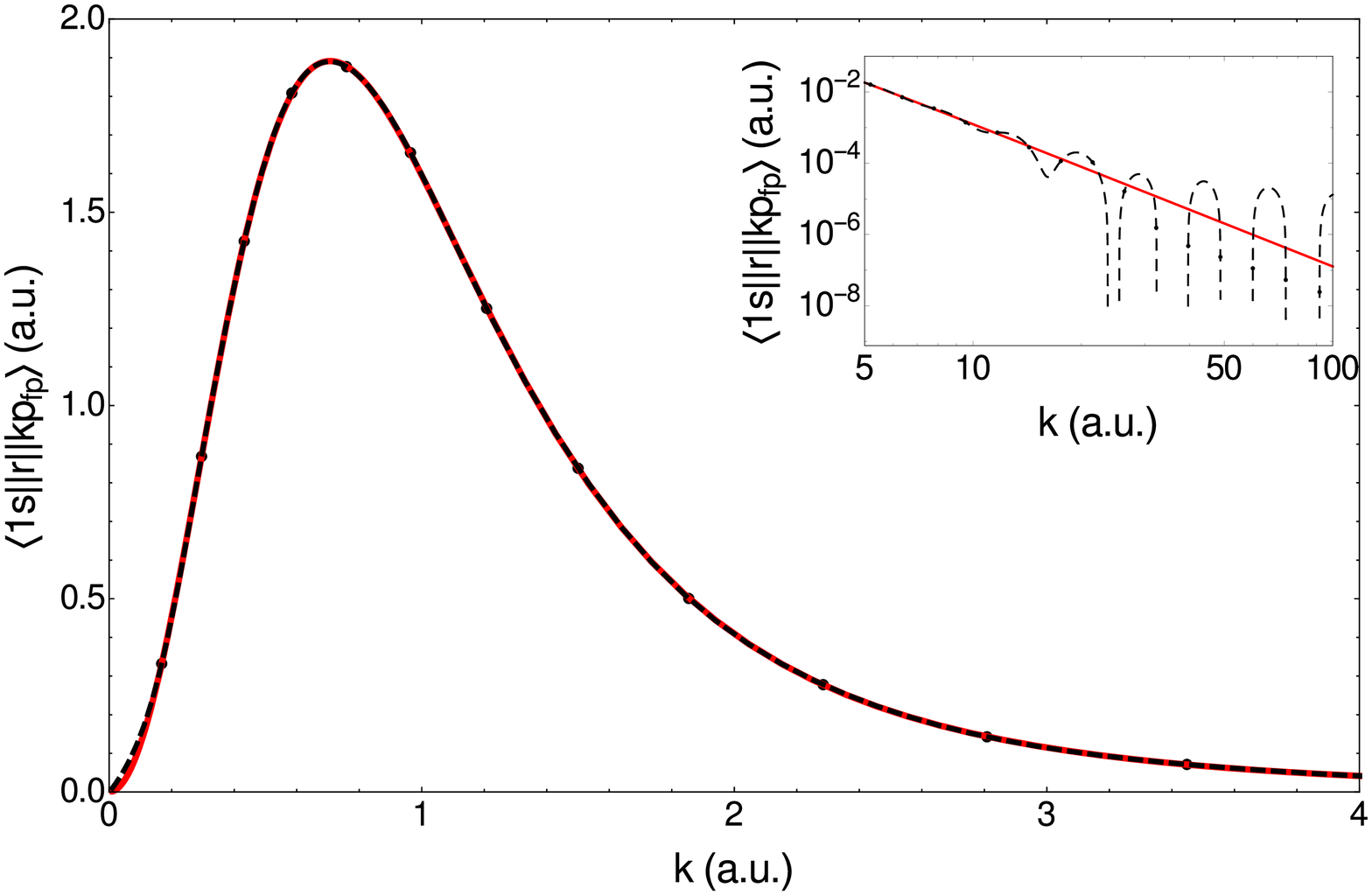} a) }
\center{\includegraphics[width=1\linewidth]{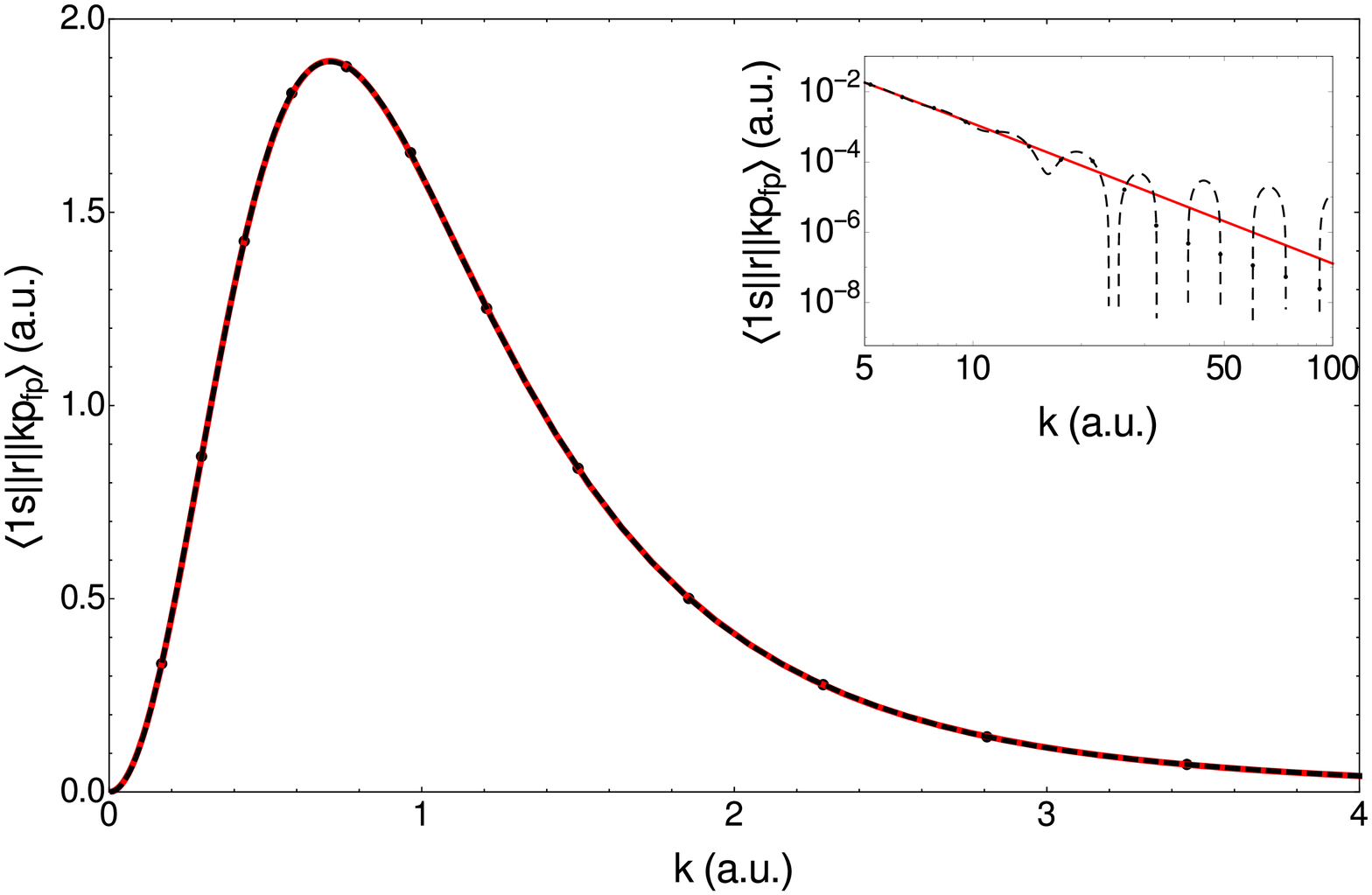} b) }
\center{\includegraphics[width=1\linewidth]{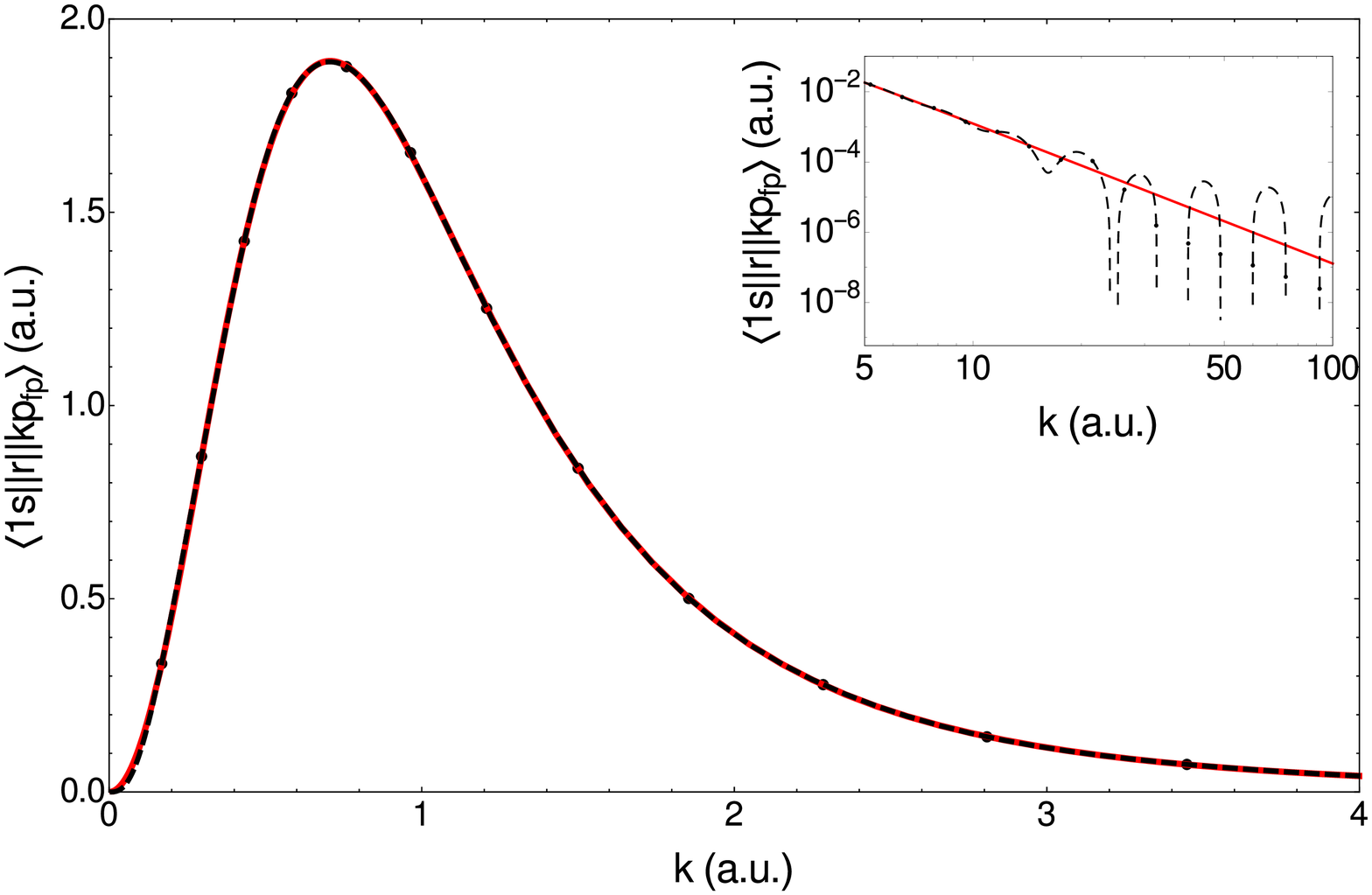} c) }
 \caption{Reduced matrix element for electric dipole transition from hydrogen ground state to free spherical wave. Solid red line represents analytic formula (\ref{AnalytFree}), dashed black line represents result of fitting using Eq. (\ref{IntpForm}) and (\ref{FreeExp}), whith correspond to Bessel-Hankel sampling function of the order a) $\nu = 1/2$, b) $3/2$, c) $5/2$. Solid black dots are matrix elements given by Eq. (\ref{HellerGen}). }
\label{FreeFit}
\end{minipage}
\end{figure*}

The method proposed in the previous section is now applied to the calculation of the matrix element defined by (\ref{AnalytFree}). In equation (\ref{MEdecomp}) we take the sampling functions $\langle m | k p_{fp}\rangle \equiv \langle m | k\rangle$ given by Eq. (\ref{SFfree}). Fig. \ref{DensityFree} represents the smooth interpolation, $s(k)$, of the set of points $\{k_n, n\}$ so that $s(k_n) = n$. Function $y(k)$ interpolates the set of points $\{k_n, y_n\}$, where $y_n$ is $n^{th}$ zero of the Bessel function of the order $\nu$, so that $y(k_n) = y_n$. Fig. \ref{SFcompare} compares the Bessel-Hankel sampling functions with $m=1$ and parameter $\nu=0.5, 1.5, 2.5$ with the exact $\langle 1 | k \rangle$ given by (\ref{AnalytSF}). One can see that those functions almost coincide exactly for $\nu=1.5$, while for  $\nu=0.5, 2.5$ there are small deviations. It is clearly seen that the zeros of the exact function (\ref{FreeExp}) only approximately coincide with eigenvalues $k_n$, while those of Bessel-Hankel sampling functions coincide {\em exactly}. Another important difference is that  the number of zeros for the exact sampling function (\ref{AnalytFree}) equals to the number of eigenvalues $N$, while Bessel-Hankel sampling functions have an infinite number of zeroes. We expect, therefore, both functions to coincide in the limit of a complete basis set with a dense distribution of eigenvalues and for the fitting to be accurate for a finite-dimensional representation except in the high-energy limit.

Fig. \ref{FreeFit} presents the result of interpolation of the reduced electric dipole matrix element $\langle 1s\|r\|k p_{fp}\rangle$ using (\ref{MEdecomp}) with Bessel-Hankel sampling functions. The fitting deviates from the exact solution as the wavenumber $k$ increases, but replicates the behavior of the exact fitting (\ref{MEdecomp}) to a very good approximation, as shown in Fig. \ref{Free}. The best results are obtained for $\nu=1.5$, which has the same asymptotic behavior as the exact p-type free particle wavefunction. One can see, however, that s-type ($\nu=1/2$) and d-type ($\nu=5/2$) fitting functions give a good fit as well. In fact the difference between s,d, and p-type fittings is noticeable only for small kinetic energies, where the sampling functions are proportional to $k^{(\nu + 1/2)}$. 
\begin{figure}[t]
\includegraphics[width=1\linewidth]{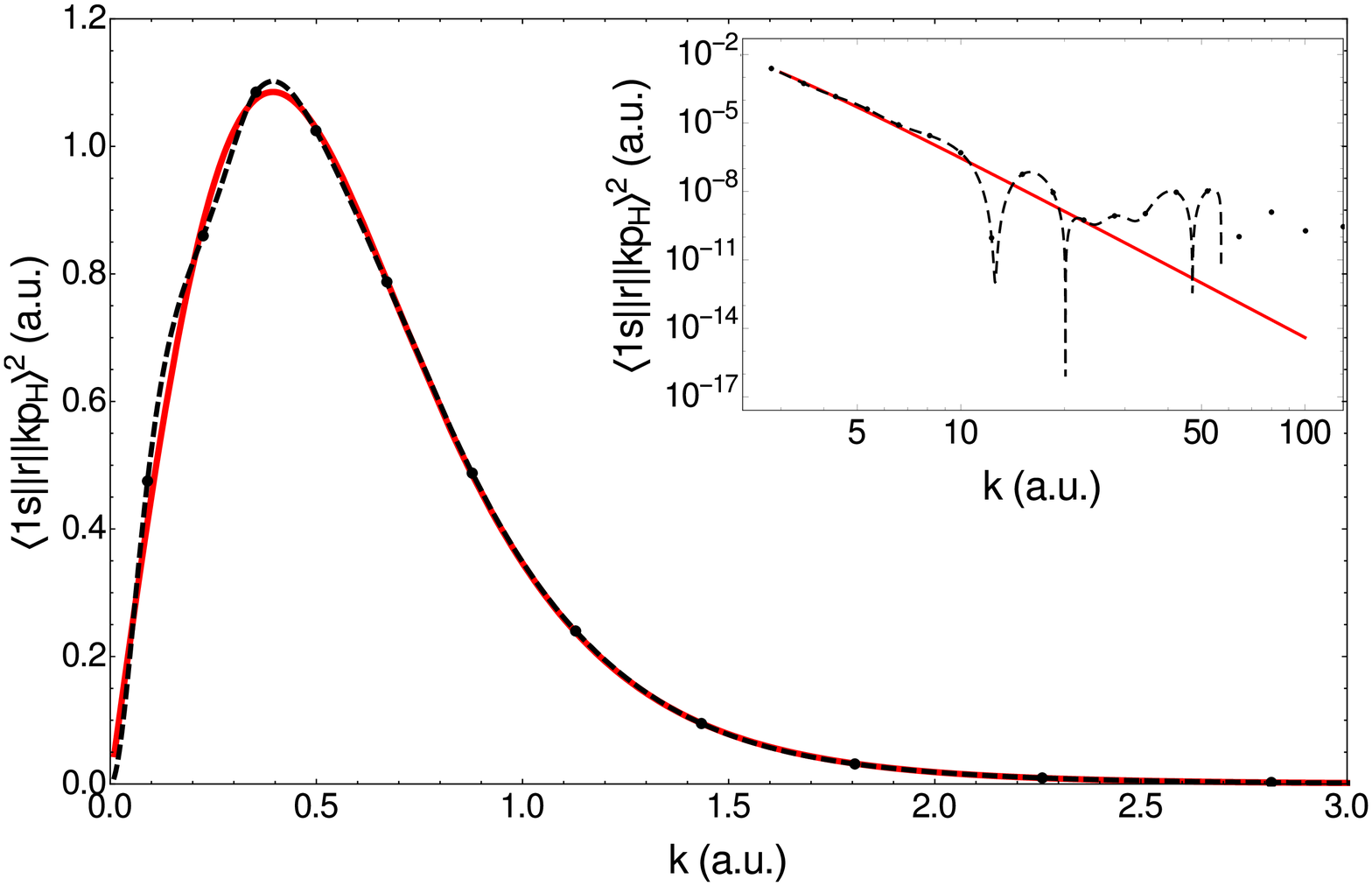} 
 \caption{Squared matrix element for electron electric dipole transition from hydrogen ground state to continuum. Solid line represents analytic formula (\ref{HAnalyt}), dashed line represents numerical result using Gaussians (\ref{HMEdecomp}). Solid black dots represent matrix elements with $k=k_m$ given by (\ref{HMEdecomp}) with a single term $\langle m | k_m p_{fp}\rangle$ included. }
\label{Hydrogen}
\end{figure}

\section{Hydrogen and Sodium photoionization}

In order to demonstrate the efficiency of the proposed method for calculation of photoionization matrix elements, we consider next the hydrogen atom. The exact expression for the reduced electron dipole matrix element is given by \cite{BetheSalpeterBook}
\begin{equation}\label{HAnalyt}
\langle 1s\|r\|k p_{H}\rangle^2 = 2^8 \frac{k}{(1+k^2)^5}\frac{e^{-{\frac{4\text{arctan}(k)}{k}}}}{1- e^{-2\pi/k}}.
\end{equation} 

To obtain the hydrogen continuum pseudo-states we used p-type even tempered Gaussian basis set with $N=35$ basis set functions with $\alpha_a = \alpha \beta^{a-1}$, $\alpha = 0.001$, and $\beta = 1.5$. The six
lowest-energy p-type states obtained using this basis set have negative energies and correspond to bound states. Therefore unlike the example described in the previous section, where all the states had positive energies, the negative-energy states are excluded from the summation in (\ref{MEdecomp}). This leads to the expression for wavepacket representation of hydrogen photoionization matrix elements given by
\begin{equation}\label{HMEdecomp}
\langle 1s\|r\|k p_{H}\rangle = \sum_{(m=1 | E_m > 0)}^N \langle 1s\|r\|m\rangle \langle m | k\rangle,
\end{equation}

\begin{figure*}[t]
\begin{minipage}[t]{0.49\linewidth}
\includegraphics[width=1\linewidth]{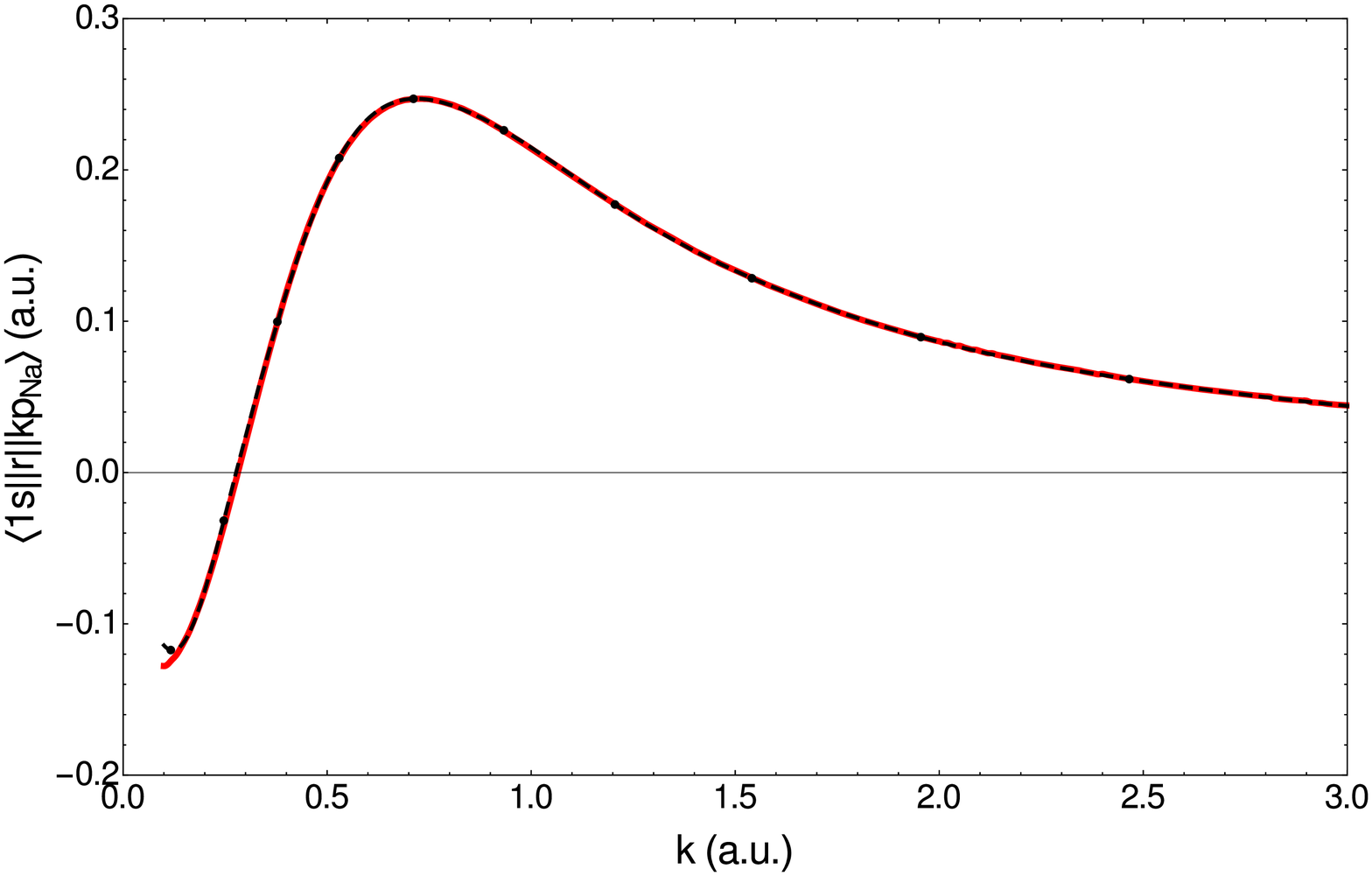}
 \caption{Matrix element for electric dipole transition of valence $3s$ electron from sodium ground state to continuum. Solid line represents finite difference solution of Hartree-Fock equations, dashed line represents numerical result using Hartree-Fock-Roothaan method with Gaussian basis set and (\ref{IntpForm}). Solid black dots represent matrix elements with $k=k_m$ given by (\ref{MEdecomp}) with a single term $\langle m | k_m p_{Na}\rangle$ included. }
\label{Sodium}
\end{minipage}
\hfill
\begin{minipage}[t]{0.49\linewidth}
\includegraphics[width=1\linewidth]{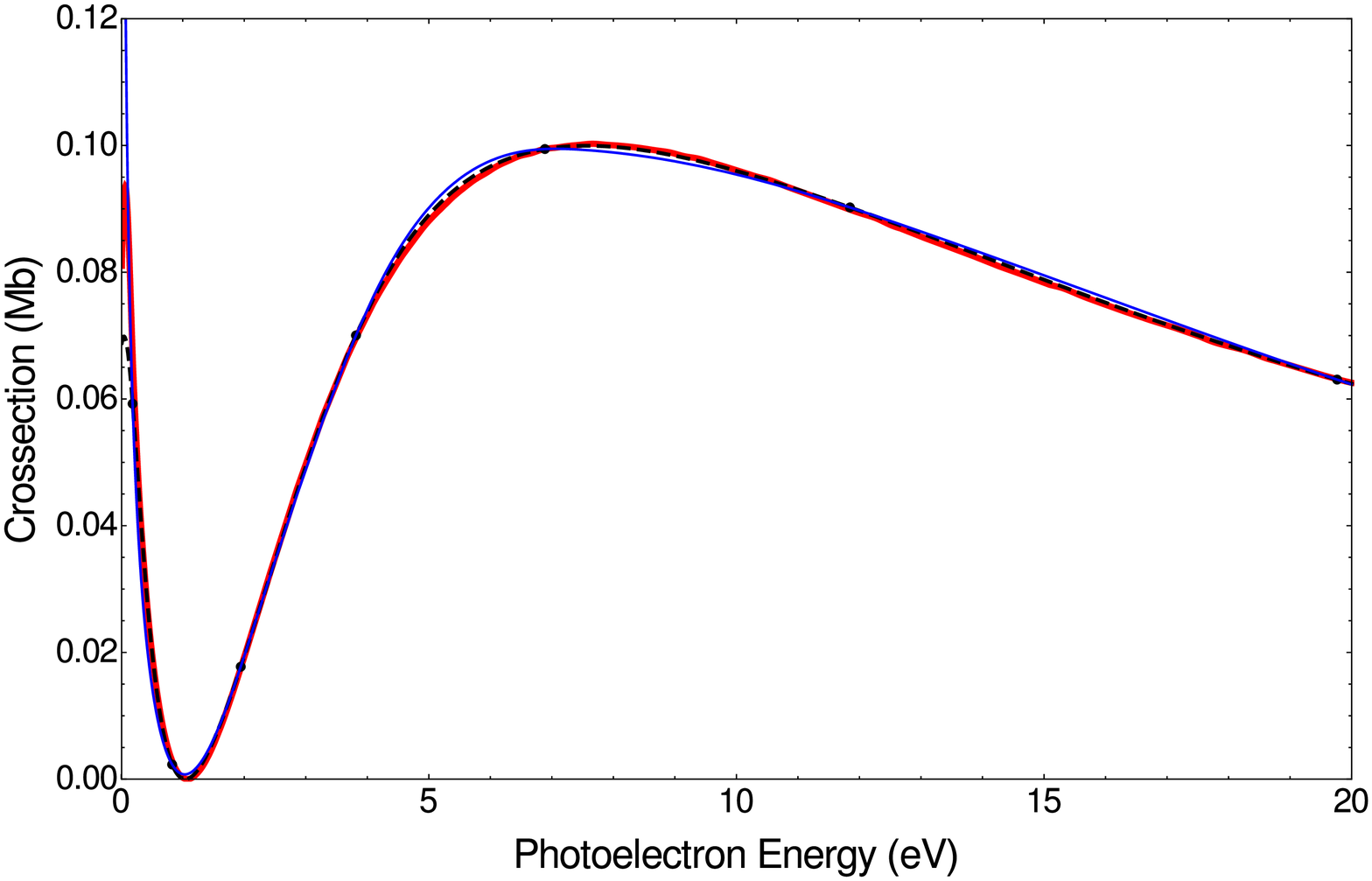}
 \caption{Sodium photoionization cross section of valence $3s$ electron as a function of photoelectron energy. Solid red line represents finite difference solution of Hartree-Fock equations, dashed black line represents numerical result using Hartree-Fock-Roothaan method with Gaussian basis set and (\ref{IntpForm}). Solid black dots represent matrix elements with $k=k_m$ given by (\ref{MEdecomp}) with a single term $\langle m | k_m p_{\text{Na}}\rangle$ included. Thin solid blue line represents interpolation black dots with third order B-splines.}
\label{SodiumCross}
\end{minipage}
\end{figure*}
\noindent where the sampling functions $\langle m | k\rangle$ are given by Eq. (\ref{SFfree}) with $\nu=1/2$, and $m=1$ denotes the lowest positive energy pseudo-state. The result for hydrogen is shown in Fig. \ref{Hydrogen}. As in the previous example of free particle final state, one can notice rapidly accumulating error of the interpolation for values of $k \approx10$ a.u. due to the reasons outlined in previous section. Also some new deviations of interpolation from an exact solution (\ref{HAnalyt}) can be observed for interpolation on the range of eigenvalues $k_1 < k < k_4$. Deviation in that region was barely noticeable in Fig. \ref{FreeFit} since the sampling function, especially with $\nu=3/2$ replicated the behavior of exact functions $\langle m | k\rangle$ to a very good approximation. Kramer theorem \cite{Kramer1957} indicates that if one would have a large number of sampling points $\langle 1s\|r\|m\rangle$ in the limit of small kinetic energy, the deviation in that region would vanish; but for finite basis sets one does not enjoy the same close match between the sampling scheme and the asymptotic behaviour of the matrix elements. For a small number of Gaussian basis functions we always have only a few pseudo-states with small eigenvalues for which $k_m \gg 1$, so the interpolation experiences some undersampling. One could expect the same thing to happen at the other end of the energy range, but Gaussian solutions tend to fail in describing rapid oscillation of continuum states $| k \rangle$ before that happens. We also performed interpolation with sampling functions (\ref{SFmodel}) with $\nu=3/2, 5/2$, but it didn't increase the quality of the interpolation. This occurs because Bessel-Hankel sampling functions with $\nu=3/2$ don't have the same asymptotic behavior near the origin as the actual sampling functions $\langle m | k\rangle$ for hydrogen (see Fig. \ref{SFcompare} b) ). 

For more complex systems with many electrons additional terms due to electron-electron interaction arise in the Hamiltonian of the system. For such systems no analytic solutions exists. Discrete states can be calculated to a very good precision by diagonalizing the Hamiltonian in a sufficiently large basis set. While the few lowest negative energy solutions are usually associated with bound state orbitals, the positive energy states are used mainly in the finite summations over virtual states that arise in perturbation theory. Using methods developed in this work we can extract some information, such as photoionization matrix elements, from these positive energy solutions, which we shall show in the case of the sodium atom. We use the simplest Hartree-Fock method for calculations of its ground state electronic structure and employ the frozen core approximation for calculations of the valence electron states (see for example \cite{Dzuba2005}). In this approximation the states of the valence electron are  calculated in the Hartree-Fock potential of the closed-shell electrons of sodium ion. Continuum pseudo-states were calculated with the same primitive Gaussian basis set as was used for hydrogen atom. Approximately 3\% accuracy in the energy of $3s$ orbital can be achieved using this method, although more sophisticated and precise approaches give considerably better precision \cite{Saha1988, Safronova1998}.

The result of interpolation on the photoionization matrix element in the length form of the dipole operator for sodium is shown in Fig. \ref{Sodium}. One can see that the fitted curve follows the finite difference solution very accurately up to the lowest energy eigenvalue. Fig. \ref{SodiumCross} represents photoionization cross-section obtained by substituting interpolated matrix element into expression (\ref{PhotoIon}). Note that the original derivative method by \citet{Heller1973} allows one to carry out interpolation of the squared matrix elements only. The results for this method are added for comparison as a thin solid blue line in Fig. \ref{SodiumCross}. Third-order B-splines were used to carry out interpolation since higher order splines give worse accuracy.

\section{Conclusions}

We have demonstrated how the Kramer sampling theorem can be used to recover photoionization matrix elements from conventional $L^2$ discretization of electronic spectra, such as one obtains from the positive energy solutions of the Roothaan equations. This procedure has being demonstrated and discussed for hydrogen and sodium, but can be easily applied to more complex atoms and molecules. The main advantage of proposed method over the spline interpolation is that it is globally defined withing the eigenvalue range of up to few hundred eV. The proposed method has a close connection with Shannon-Nyquist sampling theorem widely used in digital signal sampling and processing. Therefore some important refining techniques that are widely used in signal processing can be adopted to refine the evaluation of continuum processes in complex quantum systems.

Here, we have solely considered atomic radiative transitions, but the Gaussian basis set methods of quantum chemistry have been developed for molecular applications. Decomposition of molecular states into their atomic constituents is always possible, so that the atomic sampling methods developed here may be applied to each atomic component and convolved to produce molecular information by superimposing quantum mechanical amplitudes. The implementation of these ideas is the subject of ongoing research that will be reported in future publications.

\section*{Acknowledgements}
The authors acknowledge the support of the Australian Research Council through the Centres of Excellence for Coherent X-ray Science and Advanced Molecular Imaging. A. K. is grateful to Daniel Lewis for a useful discussion during the work on this manuscript.

\bibliography{KSQ_Ref}

\end{document}